Experiments carried out at our laboratories have led to observing a new type of electrification, now called surface photocharging. For a consistent description, we are weighing carefully the experimental evidence against the common knowledge of frictional electrification and analyze it on the premise that photocharging may be the electric face of the structural damage produced at the surface by photodesorption or laser sputtering. We find the emerging picture well consistent and elaborate photocharging mechanisms based on familiar photodesorption steps. The first mechanism based on negative-U operates in the visible and near infrared. It involves photoelectron capture by negative-U metal sites at surface dangling bonds and photohole capture by negative-U sites at active bonds. Dihole formation leading to bond breaking is the likely prerequisite of metal ion desorption. Alternatively, unihole capture at trapped electron sites can give rise to self-trapped excitons (STE) which decay leading to surface defect formation, the prerequisite of metalloid ion desorption. As a result of the coupling to normally polarized modes, surface metal or metalloid ions are desorbed while complementary ionic charges appear in the subsurface region. Alternative photoelectrification mechanisms, such as the photostimulated migration or reorientation of surface molecule ions operating in the far infrared are also discussed. Implications for the practical use of the phenomenon are also mentioned.

1. Foreword

Perhaps one of the first conscientious encounters of man with electricity dates back to the frictional electrification observations by Thales of Miletus (624-545 BC). More than 20 centuries later Faraday has asserted the dual appearance of electricity in electromagnetic and electrostatic experiments, respectively.

Apparently, it was not until the late fifties of the twentieth century that the first systematic attempts were made to put the electrification phenomena on a sound scientific basis. With reference to Harper's original publications,[1] and then to subsequent work by Kornfeld,[2] with 17 available papers in Russian journals, we pay tribute to Thales' followers in modern times.

To begin with a definition, a body is electrified when its charge neutrality is unbalanced by some means giving rise to an electrostatic field around it. In the contact process, electronic or ionic charge is exchanged between two bodies which acquire equal in magnitude but opposite in sign net charges. Upon separation of the bodies, the acquired electrification charge produces an electric field around each body. There are two basic forms of the electric field around an electrified body:
- field arising from acquired bulk, surface or linear electric charges;
- field arising from an induced electrostatic polarization.

Apart from the classic frictional means, there are other specific ways of inciting an electrification, as mentioned here and there:
- tribo-electrification (by indent, strike);
- cleavage electrification (by cleaving crystal surfaces);
- dislocation electrification (by plastic deformation);
- sono-electrification (by strong acoustic wave);
- piezo-electrification (by mechanic impact in piezoelectric material);
- pyro-electrification (by heating of pyroelectrics);
- ferro-electrification (in ferroelectrics), etc.

Experimental observations have revealed two other means of electrification, such as induced by light or ionizing radiation onto a variety of bodies. One is the polarization electrification by light that has been employed in electrophotography for some time.[3] Another is the photocharging electrification reported lately.[4] In the likely process, ionic charge photodesorbed from A-body is transferred to and adsorbed at B-body. Although both conducting and poorly conducting solids apparently electrify by photocharging, our present work will focus onto the latter category in view of a number of well-documented photoeffects, such as the photogeneration of free charges in materials with a low dark conductivity.

There has been a debate in the literature for quite some time as to the origin of experimentally observed short-circuit currents on illumination of metal surfaces.[1] However, no consensus has yet been reached on the photocharging mechanisms involved. Moreover, the controversy has deepened since similar currents have been detected on illuminating semiconductor or insulator surfaces with light on both sides of the absorption edge.[2] All these observations provide examples for a photoelectrification mechanism of a nearly universal application to the charging of solid surfaces. We use the term "photoelectrification" for designating the photocharging observed by monitoring the short-circuit currents triggered by a photon pulse across a solid.

In this paper, focusing on the poorly conducting solids we argue that photo-electrification is very likely the electric face of the photosputtering of nonmetallic surfaces. We suggest at least two photosputtering steps to operate in the metal and metalloid sublattices, respectively, of a binary compound. In the former, a surface chemical bond is broken as a negative-U dihole center forms upon it. As the bond breaks up its metal component is ejected from the surface through coupling to a normally polarized vibrational mode. In the latter, ion desorption is incited by self-trapped exciton (STE) formation and relaxation at a surface ion site. Again, an ion is ejected from the surface through the aggitation of a normally polarized vibration. Both the metal and metalloid sputtering channels involve sequences of elementary steps from the generation of photoexcited free electrons and holes to the evolution of an ionic atmosphere outside the surface.

As a result, any illuminated solid will acquire a halo of ionic atmosphere the extension of which will depend on the light intensity as well as on fundamental characteristics of that solid. The ionic halo will disappear sooner or later after switching off the illumination.

## 2. Contact or frictional electrification

Apparently, friction only serves to enhance the efficiency of surface contact by increasing the contact area per unit time (Alessandro Volta). In what follows, we shall summarize what is presently known about the mechanism.

### 2.1. Kornfeld's suggested mechanism

In the late sixties of last century, Kornfeld has suggested that each insulating body carries a space charge which is compensated for by layers of impurity adions adsorbed on its surface from the environment.[5] As two bodies of different materials are brought into contact, their respective impurity layers mix up and redistribute between the two rubbing surfaces. The individual electric balance between bulk and surface being disrupted, the two bodies are no longer neutral electrically when separated. A bulk electric charge compensated by environmental impurities is thus inherent of each material body in Kornfeld's view. He did not elaborate but mentioned the possibility that the bulk charges might be related to imperfections.

### 2.2. Space charge in ionic or nearly ionic crystals

Nevertheless, at about the time of Kornfeld's original research, it has become increasingly appreciated that charged point defects and aliovalent impurities in an ionic crystal form space charge layers in depth compensated for by an excess of host and/or impurity ions on the adjacent surface.[6] Beyond any doubt, the ionic space charge layers provide a model for Kornfeld's compensated body. As the ionic surface is rubbed by another substance, some of the compensating surface ions may be removed leading to an electric unbalance between surface and bulk which electrifies the ionic crystal. The subsurface space charge layers or related point defect structures are quantitatively described by the Poisson-Boltzmann equation whose spatially damped solutions are shown in Figure 1.[6]

Depending on the duration $\tau$ of an electrification experiment as compared to *Maxwell's relaxation time* $\tau_M = \varepsilon/4\pi\sigma$, where $\varepsilon$ is the dielectric constant and $\sigma$ is the (ionic) conductivity, there are two conceivable extreme responses of the ionic space charge layer to an external perturbation $\Delta\Sigma$ of the surface charge density $\Sigma = (k_B T \kappa \varepsilon / 2\pi e)$ sinh($\frac{1}{2} e\varphi_\infty / k_B T$) ($\kappa = r_D^{-1}$ is Debye's screening radius, $\varepsilon$ is the static dielectric constant, $\varphi_\infty$ is the bulk electrostatic potential, $e\varphi_\infty / k_B T = \ln\Psi$, $\Psi$ is the average impurity mole fraction):

(i)      $\tau \ll \tau_M$. The system is far from a charge equilibrium and the additional surface density $\Delta\Sigma$ is not compensated by the subsurface space charge $\rho$ which compensates for $\Sigma$ alone. The solid will remain electrified as long as the $\Delta\Sigma$ ions are not reabsorbed by the environment to restore the original ($\Sigma,\rho$) equilibrium state of the solid. This is the likely case of a poorly conducting solid.

(ii)      $\tau \gg \tau_M$. The system is in a new charge equilibrium ($\Sigma+\Delta\Sigma,\rho+\Delta\rho$) state where the total surface charge density $\Sigma+\Delta\Sigma$ is compensated for by a new equilibrium distribution ρ+Δρ within the subsurface space charge layer. Due to the rapid charge compensation, the solid will not be apparently electrified during the time of an experiment. This is the likely case of a good conductor.

It is now intuitively clear that for characterizing the electrification power of a solid, an estimate of Maxwell's relaxation time $\tau_M$ is essential. We make use of the simple formula $\sigma = en\mu$ for the conductivity by dominant carriers of density $n$ and mobility $\mu$. Setting $n \sim 10^{15}$ cm$^{-3}$, which is well within the spectroscopic detection limits, we get $\tau_M \sim 8\times10^{-7}/\mu$ where the mobility $\mu$ should be taken in *esu* so as to obtain $\tau_M$ in seconds. We further set $\mu = (p / 2k_BT)\, a\nu\, \exp(-U/ k_BT)$ where $p$ is the electric dipole moment, $\nu$ the attempt frequency and U the barrier for carrier migration, if any, $a$ is the lattice constant. $e$, $k_B$ and T have their usual meaning. Inserting the values typical for an ionic crystal (KCl) we get $\mu \sim 10^{-11}$ esu for $p = 1$ eÅ and U = 0.7 eV, the barrier for cation vacancy migration. Inserting we get $\tau_M \sim 8\times10^4$ sec = 22 hrs. Clearly, we have case (i) holding practically, so that KCl is electrifiable in any reasonable-time contact experiment.

Space charge layers extend around dislocation cores too which cores harbor the compensating excess ions, host or impurity. This comes to explain the origin of electrification accompanying plastic deformation of ionic or nearly ionic crystals.

### 3. Photoelectrification

Concomitant with the two general electrification means, we can expect two resulting electrified states of material bodies to be induced by light, and indeed, both have now been proven experimentally. The polarized electrified state called photoelectret has been discovered by Nadjakoff in the late thirties of last century.[7] The photocharged electrified state has been observed and investigated in his laboratory some fifty years later.[8]

### 3.1. The photoelectret state

The electret state is best modelled by a crystal containing extrinsic divalent impurity cations and cationic vacancies. These pair at sufficiently low temperatures to form impurity-vacancy (I-V) complexes. I-V complexes can reorientate along specific crystallographic directions through thermally-activated vacancy hops around the impurity. On applying an external electric field along a reorientational direction at a temperature $T_P$ within the activated range, I-V dipoles can be all alligned along that direction creating a polarized macrostate, the *'electret state'*.[3] The formation of an electret state is evidenced by the flow of polarisation currents across the sample. This state can be kept stable under field-off at temperatures $T_S \ll T_P$ well below the activated range. Clearly, one should observe depolarisation currents on going back to $T_P$ under field-off.

In a number of systems, the vacancy hopping rate requiring considerable activation energies as the impurity is in its ground electronic state becomes greatly enhanced if that impurity is raised to its excited electronic state. The physical background of photoelectrification is the photoreduced activation energy. Now the polarized macrostate named a *'photoelectret'* is again thermally stable with field-off at low enough temperatures.[3] We see that the photoelectret is an electrified state produced by the combined action of active light and aligning field upon a system of elementary electric dipoles, otherwise orientated at random. Photoelectrets apply to electrophotography as widely described elsewhere.[3]

An electrostatic polarization, viz. a photoelectret state, may also be formed in semiconductors by photoexcited charge carriers.[8]

### 3.2. The photocharged state

### 3.2.1. Photocharging experiments

Upon illuminating a solid sample with a modulated (chopped) light, a potential difference is observed to occur varying in concert with the chopping rate of the incident radiation. In a practical experiment the electric signal is being measured between an electrode pressed onto the solid and a grounded electrode.

The photocharged state has been observed almost occasionally during acousto-electric measurements: a light-modulated electric signal has been measured across a transverse-acoustoelectric-voltage structure in the absence of an acoustic wave for a sample in which no photoelectric effect has been expected to occur at the given wavelength. For investigating the photocharging effect the sample is placed inside a condenser cavity with one plate semitransparent to secure illumination of the sample. The illuminating light is disc chopped producing rectangular pulses of luminous flow. The voltage from the active plate across the illuminated cavity relative to the grounded plate is passed onto a lock-in amplifier for synchronous detection. The measured photovoltage coming from the sample falls across the input impedance of the amplifier, the order of several tens of MΩ. However, the input resistance should not modify the output signal at chopping rates superior to the reciprocal RC constant of the circuit producing a rate independent output.

At first sight, the photocharged state may have something to do with the photoelectret state of free carriers arising in semiconductors.[8] Indeed, it appears that photocharging leads to a distribution of free carriers, not electric dipoles. This is likely if excitation at bandgap energies is a prerequisite, as it appears. However, if the sample is nearly transparent to the illuminating light both its sides along the normal path of incidence will be charged nearly equally, the piece as a whole remaining neutral electrically for the outer observer. In this sense, namely, the photocharging may be regarded as a superficial electrification effect which is only activated by strongly absorbed light.

### 4. Photon desorption or sputtering of surfaces

There are both qualitative and quantitative features common of photocharging on the one hand and the photodesorption or the laser sputtering of surfaces, on the other. They suggest that photocharging may actually be providing a quantitative measure for the electric balance during desorption or sputtering. For instance, both laser sputtering and photocharging are only observed under pulsed conditions. A striking observation is the measured relaxation times in laser electrification of nonmetals which fall along a time scale inherent of ionic rather than electronic relaxation. We also found a correspondence of the number of atoms desorbed, as estimated by means of sputtering-yield data and photovoltages measured.

The laser sputtering of nonmetal surfaces is a quantal effect which involves the creation of electronic excitations.[10] Both ions and neutral atoms sputter. The ionic sputtering will undoubtedly leave the solid in an electrified macrostate, as the process transports ionic material from the subsurface onto the surface ultimately to the environment thereby altering the distribution of surface and/or bulk electric charges. Unlike it, atom sputtering would rather leave the solid in an electrically neutral state, though this would have to be specified in each particular case. In the dark, the solid readsorbes material evolved to the environment during the lasing to eventually return to its original state before sputtering.

### 4.1. Sputtering ranges

The yield *versus* laser fluence experimental dependence is exemplified in Figure 2 for sputtering $Ga^0$ neutrals from a (110) GaP surface.[9] $Ga^+$ ions are found to exhibit a similar trend. The observed intensity dependence is essentially nonlinear: We see the sputtering yield to be negligible below a threshold fluence. Increased-sensitivity yield measurements show that adsorbed atoms sputter in the lowest fluence ranges, then come host atoms at defect sites following suit in a subsequent fluence range, and finally host atoms at kink sites sputter in the highest fluence ranges before overall ablation starts.[11] It is essential that sputtering is a low-yield thermally activated process. The thermal activation suggests that the ion ultimately sputters due to its coupling to a normally polarized vibrational mode.

### 4.2. Sputtering mechanisms for poor conductors

The experimental sputtering yield *versus* laser intensity (fluence) dependence is highly instrumental in suggesting a mechanism. There have been a few proposals:

(i) The sputtering of ionic compounds may follow the steps of a Frenkel pair defect formation near the surface:[10] A pair is produced through the respective decay channel of a self-trapped exciton (STE), then the interstitial is ejected out of the surface by virtue of its coupling to a normally polarized vibrational mode. The predicted intensity dependence is supralinear.

The subsequent models all predict super- or non-linear intensity dependences.

(ii) The *Varley-Feibelman-Knotek* onsite dihole process is assumed responsible for x-ray sputtering of compound materials.[12] In its first step, creating a core hole is followed by relaxation to the valence band top. The released energy leads to the formation of an electron-hole pair via Auger process. The second hole remaining at the original hole site, a doubly (positively) charged species results which being an antisite defect is ejected electrostatically out of the crystal.

(iii) The *Itoh-Nakayama* negative-U dihole process is assumed crucial for lower yield laser sputtering when the photon energy is not enough to create an onsite dihole.[13] Instead, the dihole occurs as a result of unihole trapping at a bond site followed by dihole formation as a second hole is trapped at the same site at the expense of the lattice relaxation energy. An essential feature of second hole capture is lowering the long range Coulomb repulsive barrier of the first hole through Debye screening. This accounts for the observed yield threshold. A dihole appearing at a surface bond, it breaks that bond and ejects an atom through coupling to a normally polarized mode. The formal picture involves negative-U rate equations in which the hole trapping coefficient is a composite of diffusion and binding controlled components (cf. Part II). The binding component is temperature-dependent assuming Boltzmann's energy distribution. Specific yield *versus* fluence dependences derived theoretically at several temperatures are in concert with observed superlinear dependences for GaP, as in Figure 2.[14]

(iv) *Sumi*'s fermion model assumes Fermi's energy distribution rather than Boltzmann's distribution and Debye screening in order to derive a temperature dependence.[15] Due to Pauli's exclusion, photoholes occupy the higher conduction band levels the higher the temperature. Hot holes traverse the repulsive long range barrier more efficiently than do cold holes. The model predicts a quadratic intensity dependence.

(v) *Itoh* and *Singh* assume asymmetric exciton formation rather than carrier transport by the laser illumination followed by biexciton trapping at surface bonds with the electronic charges pointing away and the hole charges accommodating so as to break a bond via dihole formation.[16]

<p style="text-align:center">5. Concomitant photocharging mechanisms</p>

Hereby we illustrate our presumed photocharging steps for a poorly conducting material such as (110) GaP which undergoes the Itoh-Nakayama process of (iii) and exhibits the superlinear yield *versus* fluence dependence of Figure 2. At first, we refer to density functional calculations on GaAs showing that dangling bonds at surface metal atoms serve as preferred negative-U sites for trapping electrons doped from alkaline adatoms.[17] Our proposal is that surface dangling bonds at host metal ions alike are negative-U sites which immobilize *photoelectrons* and that negative-U sites for *photoholes* form along active Ga-P bonds. Dihole formation at a chemical bond would result in that bond's breaking. The combined outcome is the occurrence of metal neutrals $Ga^0$ at the surface, provided dielectrons and diholes appear at the same respective metal sites. However, $Ga^+$ ions will rather form and desorb in lieu of $Ga^0$ neutrals if dielectrons and diholes trap at bonds that are farther away than nearest neighboring.

The (110) $Ga^+$ ion photosputtering process, as sketched in Figure 3 (a), will be regarded as a fundamental though not at all the unique sequence of steps for *the photocharging through the desorption of metal ions*. Indeed, it is conceivable that the photocharging through metal ions may follow somewhat different steps for other faces of the same material or in other low-conductance materials.

The counterpart picture appears to emerge from our obtained photovoltage versus laser power dependence for a (111) GaAs surface, as shown in Figure 4. The latter dependence proving linear in the log-log plot, it may also favor the STE mechanism of (i). It might be that the (111) face configuration would favor a lower symmetry STE formation as a photohole at an As atom combines with a photoelectron at the Ga nearest neighbor. It is conceivable that the STE formation will incite the metalloid component to rotate at right angle around the [110] axis in order to accommodate the self-trapped hole, as it does in silicon oxide. This rotation will place the metalloid ion outwards along the Ga-As interconnecting line. Next, the STE nonradiative decay channel and the normally-polarized mode coupling will lead to As ion detachment out of the face. The process is sketched in Figure 3 (b). This sequence of steps appears characteristic of *the photocharging through the desorption of metalloid ions*.

Again, the process may not be unique and, indeed, note that the experiment in Figure 4 has been carried out under conditions quite different from Figure 2. One way or the other, the observed photovoltage dependence on the excitation laser power in Figure 4 is instructive. In the obtained dependence $\propto I^B$ the power B = 0.28 is close to the average of 1/3 and 1/4. However, the $\sqrt{I}$ dependence is the alternative to what has been said in Section 5 in that the unihole density accounts for a considerable percentage of the photovoltage during the light pulse.

## 6. Discussion

To recapitulate, the subsurface layer of the solid will be charged if basic ions are lost. In an ionic crystal picture the net space charge may be assigned to the vacancies left behind. Atoms and ions formed by negative-U will evolve out of the solid through their coupling to normally polarized vibrational modes. A cloud of intrinsic or impurity ions will spread out of the body in steady state with the illumination. During the dark period, the atoms desorbed will be attracted back to the body to restore its original state. A slower ionic relaxation will be observed.

As long as photoelectrification is concerned, the photoelectrets have since found wide applications in electrophotography, while the photocharges may now prove essential for characterizing surface conditions because of the correspondence between the number of atoms desorbed and the photovoltages measured.

Surprisingly, the variety of photoelectrified samples has largely surpassed the well defined crystalline and amorphous structures to show itself up among little expected substances. This suggests an universal property. Universality is the main feature which distinguishes photocharging from all other known photoeffects. One of the reasons behind it is that photocharging may actually follow different desorptive steps for different faces or materials with a similar qualitative outcome.

One way or the other, there may be a potential hazard of industrial proportions if the charging of ignitable substances under intense light is neglected. It is also conceivable that photocharging of clouds plays a part in thunderstorms, atmospheric electrification being too far from being understood comprehensively.

There seems to be a consensus nowadays that the chief agent in laser sputtering of adions and/or intrinsic ions from solid surfaces is the breaking of surface or subsurface chemical bonds. Another agent is the vibrational ejection that pushes a broken-bond ion out of the surface.

A particular bond-breaking mechanism has been elucidated in computer simulation experiments on a-CH under bombardment with low energy T.[18] In it, a tritium ion is subsequently adsorbed and rotated about a subsurface C-C bond and is ultimately accommodated between the two carbon atoms invalidating their bond. As a result, that particular C-C bond is broken, as its surface carbon component is freed because a hydrogen atom cannot sustain more than one bond at a time. A problem for the future is to simulate conditions under which a similar process is stimulated under illumination, perhaps by accommodating an excited H atom as a bond breaking agent.

Lately, another example has become known to us which is the laser sputtering of (111)-(7×7) Si surfaces.[19] Both adatoms and basic Si atoms have been found to sputter. The process has been studied carefully using an excimer laser. The sputtering yield derived correctly from the dependence of sputtering on repetitive pulsing has been investigated as function of laser fluence (superlinear) and wave length (resonant). The former supported the dihole bond-breaking mechanism, while the latter helped to identify the photoexcitation transitions as ones between a surface rest-atom dangling-bond band to the band of an unoccupied adatom back-bond orbital. By incorporating a time delay between the excimer pulse and the subsequent ionization laser pulse in photoionization spectroscopy experiments the authors have estimated the translational energy of photodetached atoms found to peak at 0.06 eV.[23] This helped to identify the nature of the vibrational ejection as a *phonon kick out process.*[20]

In principle, *any* normally polarized displacement of charge near the surface or within the bulk of a solid can be expected to lead to short-circuit currents.[21] Therefore, subsurface displacements of ions of whatever nature may contribute to the waveform simply by altering the initial conditions for sputtering.

Yet, the breaking of bonds is not the only mechanism that may lead to ionic migration at the surface. Light stimulated ion migration (LSIM) along with other photochemical processes have been known and widely discussed in the literature.[22] A specific mechanism is conceivable based on the strong coupling of the molecular ion to a rotational mode in a plane normal to the surface (see Figure 5). Now the molecular system can be excited by the absorption of FIR energy amounting to several renormalized vibrational quanta to perform a vertical transition to a higher-lying vibronic state in well "left". This optical energy is not enough to breaking a surface bond. If the next step is a horizontal tunneling transition to the neighboring vibronic well "right", corresponding to another orientational site, then the molecule performs a reorientational step. The excess energy will be given back to the lattice through the vibrational coupling between the promoting and accepting modes.[23] Light-stimulated local motions have been studied extensively as reorientations of CN⁻ molecule ions,[24] or as more meaningful rotations of off-center atomic ions in alkali halides.[25] Local rotations always bearing many important features of a translational migration, vibronic diagrams of the Figure 5 type can be extended to cover itinerancy, provided there is a strong coupling to an appropriate translational mode. However, in specific cases migration can be stimulated by local heating if the promoting-mode-to-accepting-mode coupling is not that strong so as to drain the excess energy fast enough.[26]

One way or the other, LSIM processes will be expected to result in short-circuit currents and therefore in photocharging under far infrared light. This makes LSIM a specific mechanism different from sputtering, mostly active in the VIS range.

Perhaps one of the most appealing occurrences of photodesorption or sputtering and LSIM is their possible involvement in multiple or avalange lightning. The modern understanding of atmospheric electrification has not reached far enough, yet there is a convention that under local convection currents $H^+$ and $OH^-$ ions become vertically separated giving rise to enormously high voltages. The local reorientation of $OH^-$ molecular ions and their interaction with a far infrared light field is well known,[27] less known is their sputtering or desorption in a visible light field. Nevertheless, it is conceivable that with sputtering at hand, there may be an avalanche lightning process triggered by an initial accidental lightning. If so, taking photocharging into account will undoubtedly exert a marked effect on our understanding of the atmospheric processes in everyday life.

Another important research area is on the photocharging of spaceship surfaces which if not reckoned with could bring signal interference or even serious surface damage through discharge. One of the earlier investigations has confirmed the accumulation of considerable photocharge ($\sim 10$ $\mu C/m^2$) on illuminating a fused silica surface by ultraviolet light in vacuo ($10^{-5}$ mm Hg).[28] Among other things, it suggests that solar radiation could be harmful to spaceships unless protective covers by poorly electrifiable materials are incorporated. In its turn, the latter stresses the importance of laboratory photoelectrification research.

Undoubtedly, the photocharging or photoelectrification effect can be considered to be an extension to the lower energy photon range of photoelectric processes occuring under ionizing radiation.[29] We conceive different mechanisms to be operative in different spectral energy ranges along the electromagnetic scale, such as vibronic reorientation or hopping migration of molecule ions in the FIR range, through photosputtering or photodesorption of surface ions at energies below the bandgap, through photovoltaic effects at or above the bandgap, through defect formation via nonradiative decay of self-trapped excitons in the ultraviolet range. The STE decay processes can be viewed as a link between the photon ranges and the ionizing ranges (x-ray and then $\gamma$-ray in the increasing quantum energy order) where direct knock-out processes may supersede.

We see that investigations in each of the spectral ranges bring heuristic or practical advantages of their own. For instance, the visible (mostly sputtering) range brings about the option of a genuine phototopography of surfaces through a 2-D distribution of photocharges (voltages) depending on the defect structure of that surface. A low transversal diffusion rate of the sputtering species apparently contributes to the contrast of a specific topographic pictire. Yet the options provided by a diversity of operating mechanisms make the photoelectrification flexible and almost universal by apllying to a very wide range of solid materials.

Acknowledgement

It is a pleasure to thank Professor D. Nenow (Sofia) for enlightening discussions and critical reading of the manuscript, Dr. A. Buroff for encouraging remarks and Dr. A. Gochev (Columbus) for recent information.

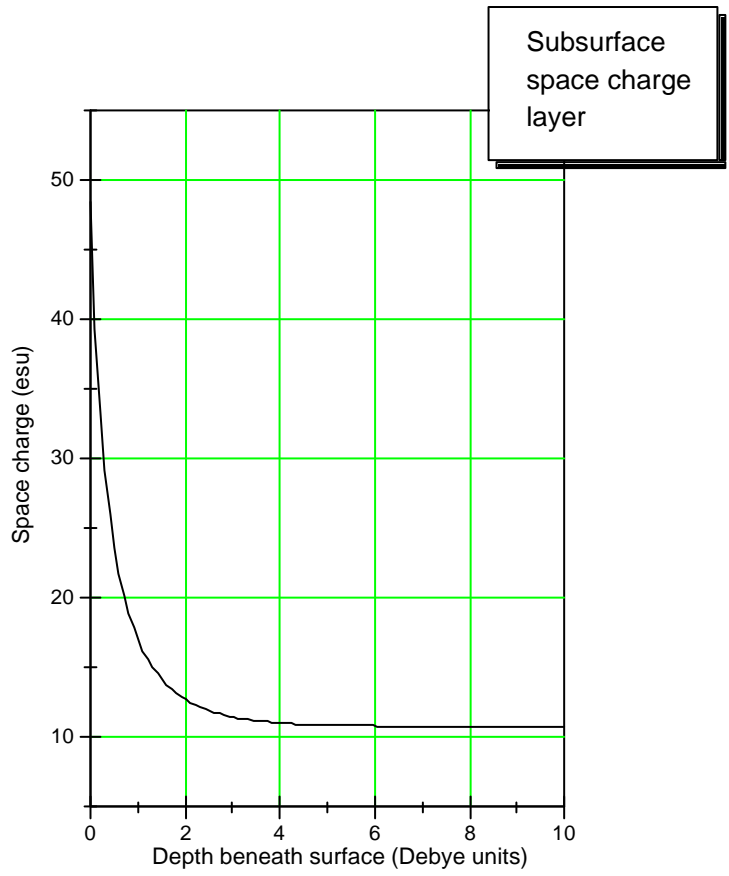

Figure 1. In-depth distribution of the ionic space charge $\rho(x)$ in an alkali halide crystal doped with divalent cationic impurities. The bulk charge $\rho$ produced by impurities and associated cationic vacancies is compensated by a surface charge density $\Sigma$ of the opposite sign. ( In the particular case $\Sigma = -7.85 \times 10^{10}$ esu /cm$^2$ at an average impurity mole fraction of $\Psi = \exp(e\varphi_\infty / k_B T) = 10^{-6}$.) The structure is a double layer confined to the crystal which remains in an unelectrified state. In an electrification experiment impurity charge density $\Delta\Sigma$ is added to or removed from the surface. Depending on the duration $\tau$ of an electrification experiment relative to Maxwell's relaxation time $\tau_M$, (a) at $\tau \ll \tau_M$ there is no time for a bulk response $\Delta\rho$ to match $\Delta\Sigma$ and the crystal will be electrified, (b) at $\tau \gg \tau_M$ there is time for a bulk change $\Delta\rho$ to match $\Delta\Sigma$ and the crystal will not be electrified.

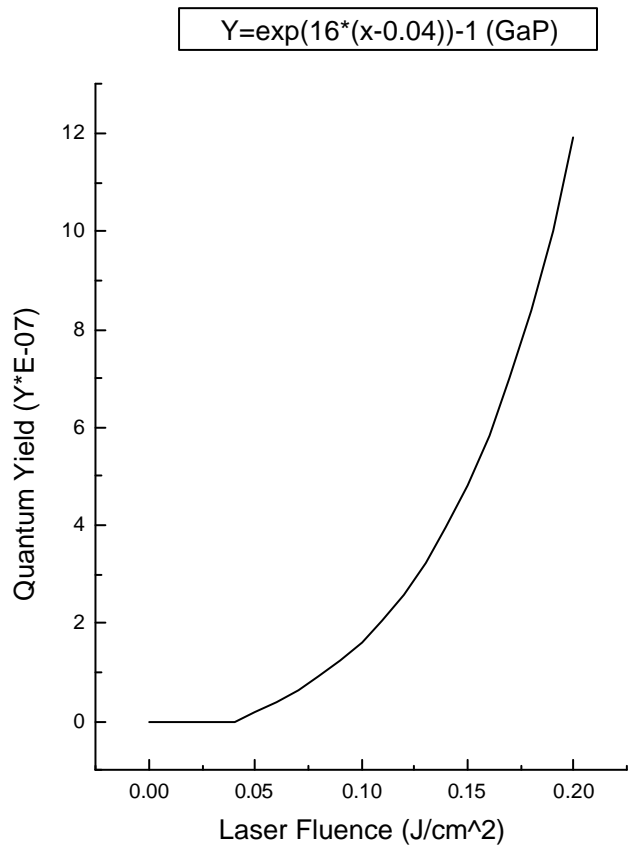

Figure 2. Typical sputtering yield Y *versus* laser fluence F dependence obtained for the emission of Ga neutrals from a (110) GaP face, as cited by Reference [13]. Single laser shots of 15 ns duration at 600 nm have been used. The Y → F dependence can be recalculated to surface photo-charge Q→F or voltage V→F data: $V = Q/C = 2eYc_\phi F/C$ yielding V=100 μV per Y=1 at F=1 J/cm$^2$ for C=1 pF.

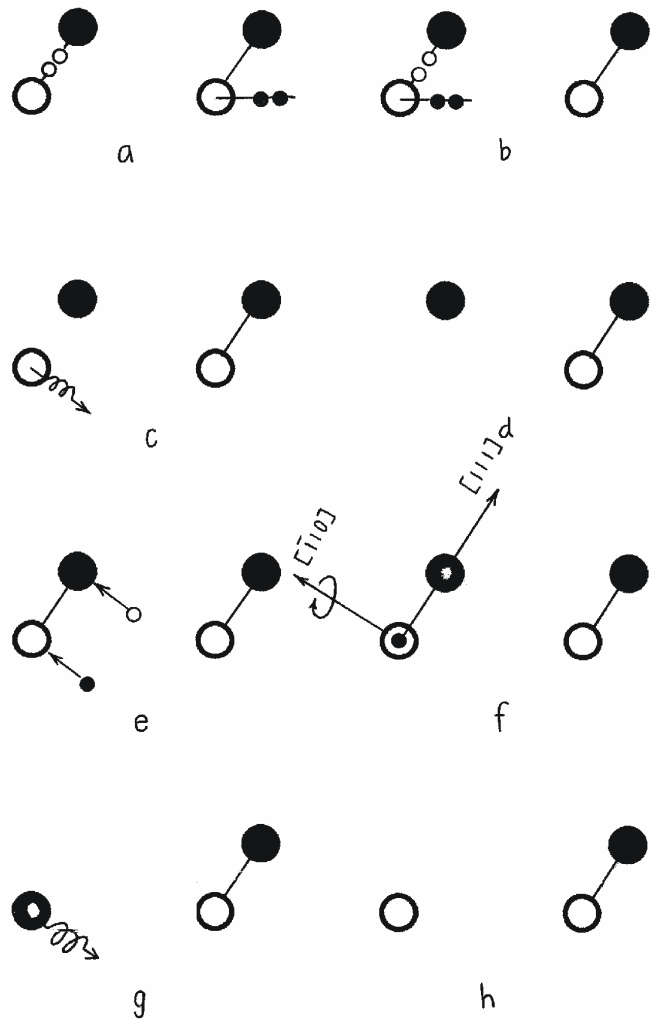

Figure 3. Illustrating the presumed elementary steps in photocharging of GaAs. Steps (a) through (d) show the negative-U desorption process for metal particles, ions in (a) and neutrals in (b): (a),(b) – dihole at active bond (left), dielectron at dangling bond (right) for separate molecules (a) and same molecule (b); corresponding spring ejection of metal ion in (c) and metal atom in (d). Steps (e) through (h) show the STE desorption process for metalloid particles. (e) – electron-hole capture at molecule (left), (f) – STE rotation normal to surface, (g) – normal spring coupling, (h) – metalloid atom ejected from the surface.

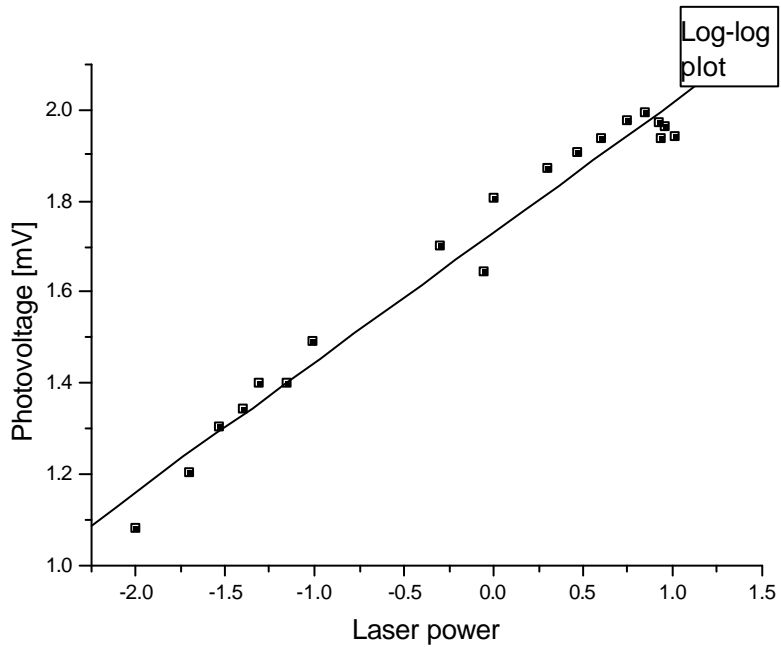

Figure 4 An experimental photovoltage *versus* laser power dependence obtained on photocharging a (111) GaAs face. The He-Ne laser at 632.8 nm shining a $\phi$2 mm spot was used chopped at a 680 s$^{-1}$ rate so as to generate a square waveform. The light intensity was calibrated through polarization filters. Plotted voltages were subject to a 20 dB amplification. A linear regression analysis of the log-log plot yields log$V$ = A + Blog$I$ with A = 1.72958, B = 0.28437. (A linear log-log plot implies $V = 10^A I^B$.) The linear regression coefficient is R = 0.98631, while the standard deviation is 0.05.

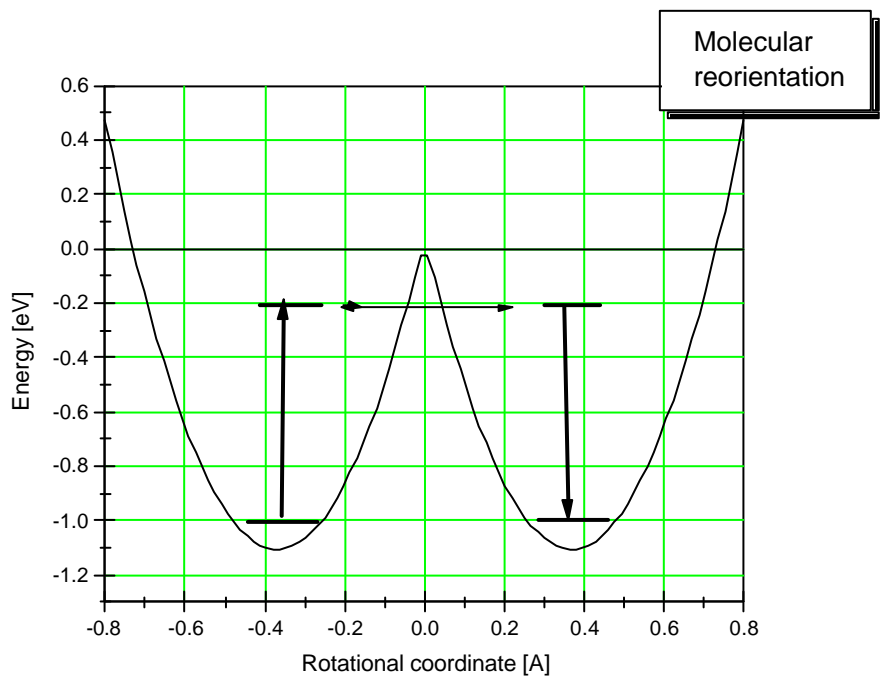

Figure 5. Vibronic energy diagram of a molecular ion along the coordinate of a rotational mode in a plane normal to the surface. Well "left" is the energy diagram of the molecule in a rotational state 1, well "right" is the energy in the neighboring rotational state 2. The molecule undergoes an optical excitation in state 1 depicted by the upward vertical arrow. A reorientational step through horizontal tunneling follows which brings the molecule from excited state 1 to excited state 2. The downward vertical arrow in well "right" marks the transition to ground state 2 as the excess optical energy is spilled through intermode vibrational coupling.